\documentstyle[11pt,newpasp,epsf,twoside]{article}
\def\edcomment#1{\iffalse\marginpar{\raggedright\sl#1\/}\else\relax\fi}
\reversemarginpar

\def\kms{\relax \ifmmode {\,\rm km\,s}^{-1}\else \,km\,s$^{-1}$\fi}
\def\ha{\relax \ifmmode {\rm H}\alpha\else H$\alpha$\fi}
\def\hb{\relax \ifmmode {\rm H}\beta\else H$\beta$\fi}
\def\hi{\relax \ifmmode {\rm H\,{\sc i}}\else H\,{\sc i}\fi}
\def\hii{\relax \ifmmode {\rm H\,{\sc ii}}\else H\,{\sc ii}\fi}
\def\h2{\relax \ifmmode {\rm H}_2\else H$_2$\fi}

\def\deg{\hbox{$^\circ$}}
\def\min{\hbox{$^\prime$}}
\def\sec{\hbox{$^{\prime\prime}$}}
\def\fdg{\hbox{$.\!\!^\circ$}}

\def\farcm{\hbox{$.\mkern-4mu^\prime$}}
\def\farcs{\hbox{$.\!\!^{\prime\prime}$}}
\def\degd#1.#2{ #1\fdg#2 }                 
\def\mind#1.#2{ #1\farcm#2 }               
\def\secd#1.#2{ #1\farcs#2 }               

\def\plottwo#1#2{\centering \leavevmode
\epsfxsize=.45\textwidth \epsfbox{#1} \hfil
\epsfxsize=.45\textwidth \epsfbox{#2}}

\begin{document}
\title{Kinematics in the Central Kiloparsec of Spiral Galaxies}
\author{Johan H. Knapen}
\affil{Isaac Newton Group of Telescopes, Apartado 321, 
        E-38700 Santa Cruz de La Palma, Spain, and
University of Hertfordshire, Department 
of Physical Sciences, Hatfield, Herts AL10 9AB, UK}

\begin{abstract}

Results from kinematic observations of the central regions of spiral
galaxies are reviewed, with particular emphasis on starburst and AGN
hosts. While morphological studies lead to important insight, a more
complete understanding of the physical processes that drive the
evolution of the central regions can be achieved with measurements of
the kinematics of gas and stars. Here, a variety of observational
techniques at different wavelengths is critically discussed, and
specific areas of interest are highlighted, such as inflow in barred
galaxies and the origin of nuclear spiral arms. A brief discussion of a
number of case studies is presented to illustrate recent progress.

\end{abstract}

\section{Introduction}

The study of the central regions of galaxies, and especially those
hosting (circum)nuclear starbursts or AGN which this conference is
focused on, is of great importance for our understanding of a variety of
astrophysical processes, ranging from those governing the existence and
sustainment of the central activity, to the dynamics and evolution of
galaxy discs. While progress on the observational front has been
arguably lacking behind theory and modelling in the areas of, e.g., bar
dynamics and fuelling of central activity, the availability of imaging
at high angular resolution from the ground and from space has led to
considerable progress on many morphological aspects (e.g., review by van
der Marel 2001; see also various papers in these proceedings). Tests of
results and predictions from theory and modelling by fully sampled
kinematic observations at similarly high resolution, however, are
forthcoming only now. The fact that for many years after the pioneering
papers on gas flows in bars (e.g., Sanders \& Huntley 1976) the only
two-dimensional kinematic mapping available was from \hi\ studies, at
angular resolutions of 15\sec\ at best, illustrates this point. Only
much later could gas kinematics be measured from molecular line
observations, and even now CO interferometers are only just breaking the
barrier of 1\sec\ resolution (see Sect.~2.4).

This review is not meant to do justice to all those excellent papers on
the topic which have appeared in the literature in recent
years. Instead, an overview will be given of what can be learned from
kinematic observations in a variety of objects, and what techniques can
be used for which purpose. Recent progress is illustrated with a
description of case studies of the core regions of two galaxies with
circumnuclear ring-like structure, M100 and NGC~5248, for which
kinematic observations lead to important tests and extensions of results
obtained from imaging and modeling. Finally, some thoughts are given on
the future of the field, which is very promising as new observatories
and instruments come on-line.

\section{Observational techniques}

\subsection{General remarks}

\begin{figure}
\plotone{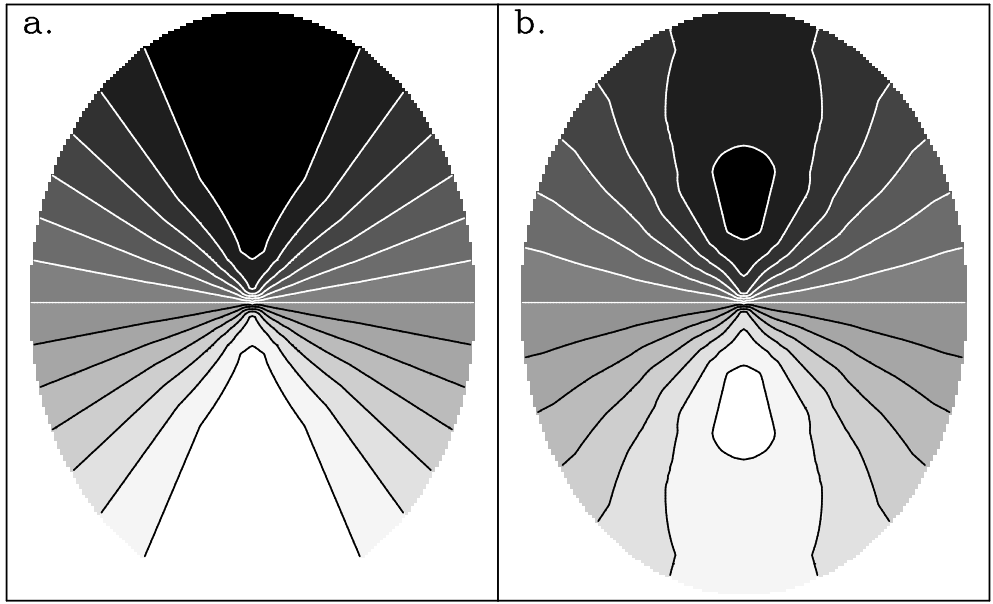}
\caption{Examples of ``spider diagrams'': velocity fields produced from
rapidly rising synthetic rotation curves, which (a.) reach a maximum and
then remain flat (i.e. constant rotational velocity with increasing
radius), and (b.) fall off after having reached a local maximum. Disk
inclination is 45\deg\ in both cases.}
\end{figure}

When making kinematic observations of galaxies, except possibly our own,
one invariably measures blue- or redshifts of emission or absorption
lines. One is also always limited to observing the line-of-sight
velocity, or the component of the true velocity projected along the
sightline. A conventional way to display kinematic results is in the
form of a velocity field, an image where (isovelocity) contours and/or
colour shades indicate the measured line-of-sight velocity. In the case
of an inclined uniformly rotating disk, the velocity field takes the
shape of a so-called ``spider diagram'', examples of which are shown
in Fig.~1. Velocity fields can be fitted to derive a rotation curve,
which is the run of (deprojected) circular velocity with radius in a
galaxy. Rotation curves often rise steeply in the central regions of
galaxies. They flatten off close to the nucleus, sometimes after a few
to tens of pc, although the measurement of this turnover depends
critically on the spatial resolution of the observations. In many
galaxies, rotation curves remain flat to radii well beyond the edge of
the optical disk, which implies that non-visible, dark, matter is
plentiful in the outer regions of those galaxies (see the review by
Sofue \& Rubin 2001 for an extensive discussion of rotation curves).

Gas and stars in a galaxy do not exclusively move on circular orbits,
and deviations from disk rotation are seen in many velocity fields. The
most important of such deviations are those caused by warps (not
relevant for the present review), by bars and triaxial potentials, and
by streaming across spiral arms (Fig.~3,~4). Other effects, due to,
e.g., in- and outflows and shocks, are harder to recognise in velocity
fields, and can only be studied by analysing and/or modeling the details
of the full data set or data cube (see, e.g., the analysis of bar
streaming presented in Knapen et al. 2000a). In many cases, a simple
axisymmetric model velocity field produced from the rotation curve
(Fig.~1) can be subtracted from the observed velocity field to highlight
deviations from circular motion, such as those due to bars or spiral
arms.

It is sometimes instructive to produce cuts through a kinematic data
set, and in fact this is often the only way to present subtle effects
seen in the data set. Such position-velocity (PV) diagrams can be made
at random positions and position angles in the data cube, and thus mimic
the result of optical/near-infrared (NIR) long-slit spectroscopy (see
below).

\subsection{Radio wavelengths -- H{\sc i} 21cm  line}

Because atomic hydrogen is rather optically thin, abundant in disk
galaxies, and traces cold atomic gas, the H{\sc i} 21 cm emission line
is a near-perfect diagnostic for disks of galaxies.  Unfortunately,
however, H{\sc i} in emission is generally not useful as a diagnostic
in the central kpc regions of galaxies, because the low column density
of the gas does not allow one to observe it at the required high
spatial resolution.

The main use of H{\sc i} in this area is to observe it in absorption
against a bright continuum source. Here, not the column density of the
atomic hydrogen, but the brightness of the background continuum source
is the main limiting factor. This brightness may be so high as to
allow the use of very large baseline interferometry (VLBI), and thus
of very high spatial and spectral resolution observations of the cold
gaseous component. This is nicely illustrated in recent work by
Mundell et al. (2001) who observe H{\sc i} at a spatial resolution of
25 mas or 1.6~pc in the nucleus of NGC 4151, and place constraints on
the location of the AGN in the nuclear region. Morganti et al. (these
proceedings, p. 000) give an overview of recent H{\sc i} absorption
results in a sample of radio galaxies, finding evidence for a wide
range of phenomena related to the AGN, such as absorbing tori or
outflows. Whereas H{\sc i} absorption is obviously a very powerful
technique, it is limited by the fact that not all galaxies have bright
enough continuum emission to allow this technique to be used to its
full extent. So although it can produce unique results at extremely
high spatial and spectral resolution, this will only be the case for a
limited numbers of galaxies which are biased by having strong nuclear
radio continuum emission.

\subsection{Radio wavelengths -- masers}

Strong line emission by masers, mostly of the OH or H$_2$O molecules,
also allows investigation of the central kpc kinematics at extremely high
resolution, using VLBI techniques. Maser emission traces high
temperatures and gas densities, and often traces rather extreme star
formation (see theoretical reviews by Elitzur 2001 and Watson 2001 for
more background).

As an example, we mention the MERLIN study of the 18-cm OH maser
emission associated with the active (AGN and starburst) core of the
ultraluminous infrared galaxy Markarian 273 by Yates et al. (2000). The
brightest of three distinct regions of radio emission, two of which have
clear NIR counterparts, harbours a 100~pc double-peaked structure, seen
in NIR adaptive optics and radio continuum images (Knapen et
al. 1997). The brightest component of the maser emission spatially
resolved by Yates et al. shows ordered motion within the central 100~pc
of the galaxy, which is aligned with the axis of the double-peaked
structure, and which thus constrains the origin of this double
peak. Other examples of the use of maser emission in the study of the
central kpc of galaxies are given in the papers by Peck et al. and
Pihlstr\"om, Conway \& Booth (these proceedings, p. 000 and 000).

\subsection{Millimeter wavelengths -- molecular lines}

Molecular line transitions in the millimeter wavelength domain,
especially those of the CO molecule, have been used for many years as
kinematic and morphological tracers of molecular gas in the central
regions of galaxies (e.g., review by Combes 1999). CO emission is often
directly interpreted as a measure of the mass of molecular hydrogen,
through the use of a conversion factor, the so-called $X$-factor. This
factor may well not be constant and its use in the central regions of
galaxies with increased pressures, temperatures, radiation fields, and
gas column densities may not be appropriate at all (e.g., Wall et
al. 1993; Combes 1999; Regan 2000).  Morphological information from CO
observations is rather difficult to interpret, not only because hydrogen
gas masses cannot be accurately estimated, but also because local
processes, e.g., star formation, may change the $X$-factor within a
small area.

Millimeter interferometers, such as ATCA, BIMA, IRAM, NRO, or OVRO, now
routinely deliver imaging at millimeter wavelengths at spatial
resolutions around or below one arcsecond. Many examples of such
observations were presented at this conference (papers in Section 000
of these proceedings). The datasets contain information on kinematics as
well as morphology. Here, the $X$-factor is generally less of a problem,
because even if the CO emission at a certain point does not quite trace
the mass of molecular hydrogen, it will most probably still trace the
same line-of-sight velocity, namely that of the molecular cloud of which
both CO and H$_2$ are part. Molecular interferometry is a field with a
tremendous future promise with the Atacama Large Millimeter Array (ALMA)
project now well underway.

\subsection{Optical/NIR: long-slit spectroscopy}

The technique of long-slit spectroscopy is often used for kinematic
measurements to derive line-of-sight velocities and velocity dispersions
along certain position angles in extended objects. Applications include
the determination of central mass concentrations and MBH masses, and
the overall kinematics of bars.

\begin{figure}
\plottwo{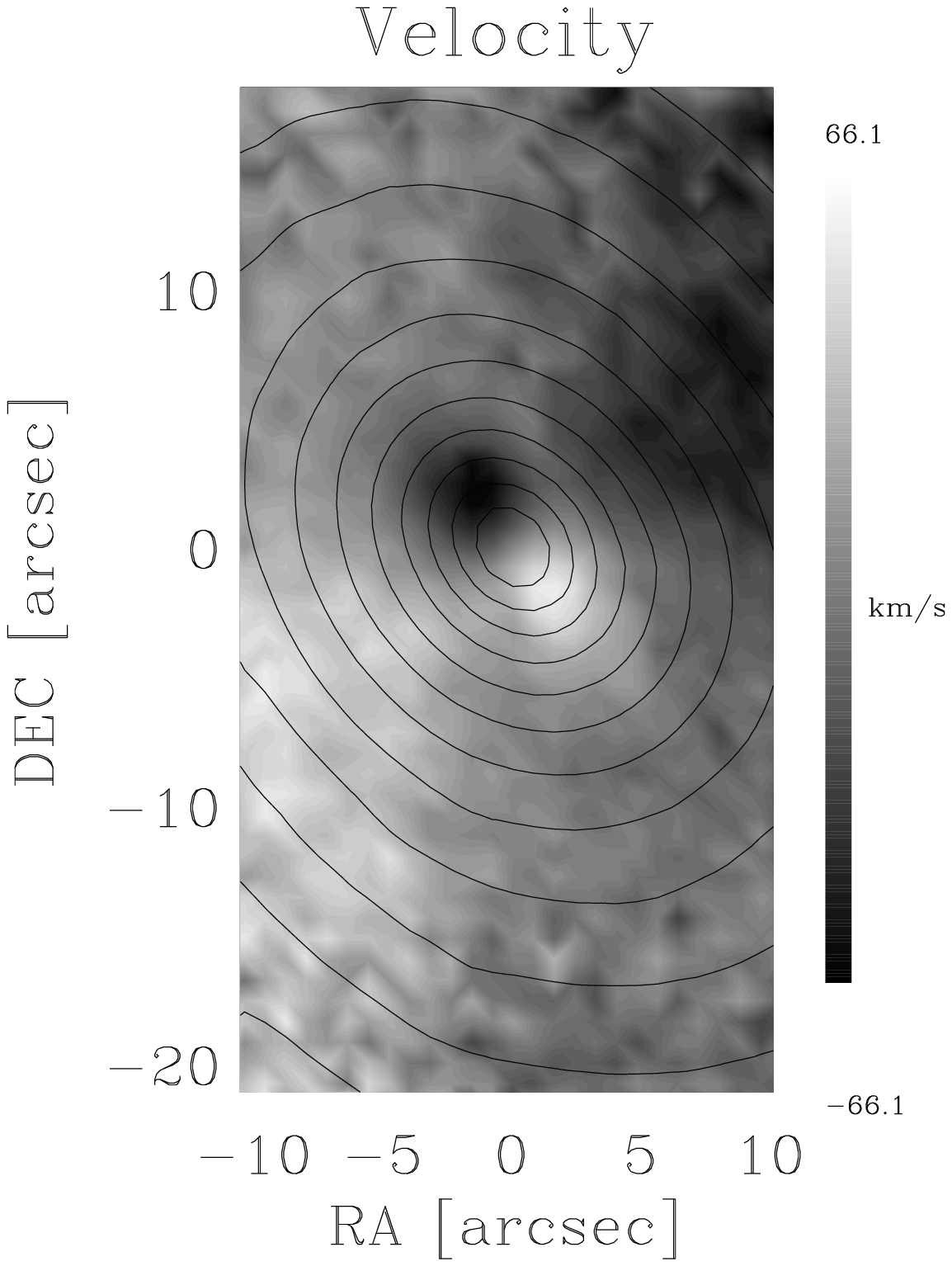}{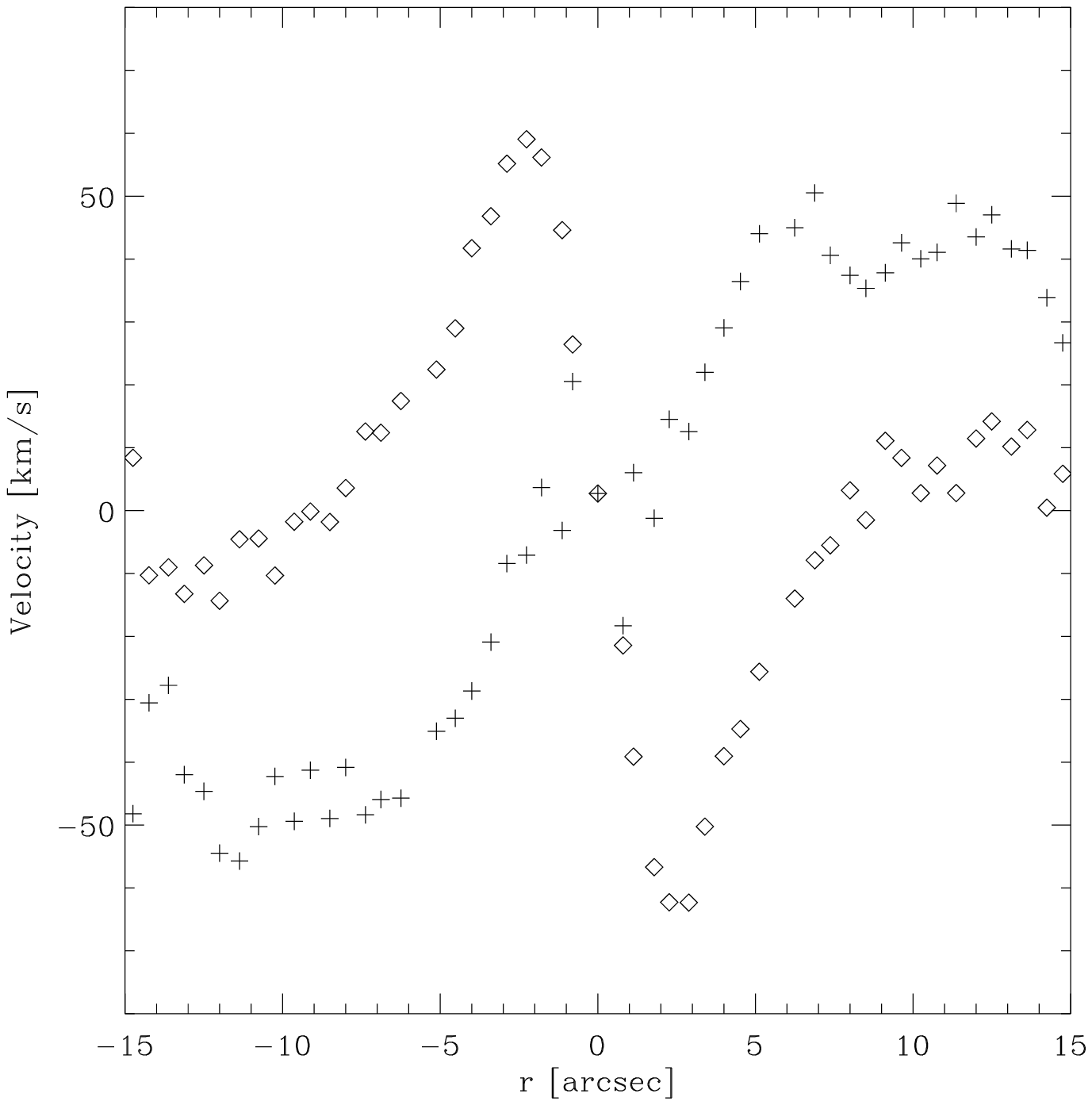}
\caption{{\it Left:} Velocity field of the central region of NGC~4365
(greys), and total intensity map (contours). The velocities were
measured using the spectral region around the Mg\,$b$ feature at
$\sim$5180~\AA. {\it Right}: extractions from the Sauron data-cube along
the major (open diamonds) and minor axis (plus signs). Based upon SAURON
data from Davies et al. (2001).}
\end{figure}

Although long-slit spectroscopy can be used to measure the kinematic
structure of the central kpc regions of galaxies, two-dimensional
velocity maps allow the precise location of different kinematic
components. This is nicely illustrated by the case of the central region
of the galaxy NGC~4365, of which Davies et al. (2001) publish the
velocity field as observed with the SAURON instrument on the William
Herschel Telescope (WHT) (Bacon et al. 2001a; see also Sect.~2.7). Surma
\& Bender (1995) obtained long-slit spectra of this same galaxy, and
interpreted the kinematic behaviour as evidence for a decoupled
core. The velocity field of Davies et al. confirms this result, but also
shows in detail how the line of nodes of the velocity field rotates by
$\sim100$ degrees in the core region of the galaxy.  Moreover, the
information on line strengths extracted from their observations allows
Davies et al. to conclude that the two kinematically distinct components
have a common age and star formation history. In Fig.~2, the SAURON
velocity field of the inner region of NGC~4365 is shown in greyscales,
with contours indicating the intensity as reconstructed from the SAURON
data (left panel). The right panel shows two synthetic ``long-slit
spectra'', extractions from the Sauron data-cube along the major axis
(open diamonds) and along the minor axis (plus signs). The difference
between these is striking but its origin can be easily understood from
the velocity field.

\subsection{Optical/NIR: Fabry-P\'erot}

The way forward in the kinematic study of the central regions of
galaxies is by mapping the two-dimensional velocity structure, or
three-dimensional imaging as this is sometimes referred to (where the
third dimension is velocity, the third axis in cubes of images). Radio
interferometry gives this kind of imaging, but also in the optical-NIR
there are several ways to achieve this. One of those is by the use of
Fabry-P\'erot interferometers, which, after data reduction, produce a
set of fully-sampled images, each at a slightly different wavelength or
velocity. These data sets or data cubes can then be treated in several
ways, but much in the same way as radio interferometry data sets are
analysed, e.g. through Gaussian or moment fitting of the individual
spectral profiles.

The advantage of Fabry-P\'erot interferometers is that they deliver
fully sampled high spatial resolution kinematic data sets over decent
fields (few arcmin at subarcsec resolution). Data reduction used to be
cumbersome but is not a serious problem anymore with modern computers
and data analysis software. A disadvantage of this technique is that it
is rather expensive in observing time because only a very small
wavelength range can be scanned during one observation (almost always
one spectral line only). Fabry-P\'erot interferometers remain powerful
for kinematic studies of galaxy disks and the central regions thereof
(Fig.~3, 4), but for small fields of view in central kpc regions
integral field spectrographs (IFSs) are very competitive, as described
briefly in the following section.

\subsection{Optical/NIR: Integral field spectrographs}

IFSs are becoming more and more popular, and nowadays form part of the
standard instrumentation suite of most major telescopes, including those
in the 8-m class. The basic idea is to divide the incoming focal plane
image into a number of small independent sub-images, sometimes referred
to as micropupils.  These are dispersed and form individual spectra,
which are subsequently captured on a CCD or NIR array detector. Spectra
are thus obtained simultaneously for a range of $(x,y)$ positions across
the image. In the data reduction stage, the individual spectra can be
combined into maps of the source by extracting specific parameters, such
as velocity, line strength, or velocity dispersion of a certain spectral
line. In one observation, one can thus obtain maps of all these
quantities, built up of a number of independent ``pixels'' which equals
the number of individual micropupils into which the original focal plane
image of the source was divided. This basic idea is common to IFSs,
although in practice several techniques are in use to divide the focal
plane image, namely by employing a lenslet array (e.g., SAURON, Bacon et
al. 2001a; Krajnovic, these proceedings, p. 000), a fiber bundle (e.g.,
INTEGRAL, Arribas et al. 1998; Garc\'\i a-Lorenzo, these proceedings,
p. 000), or an image slicer (e.g. the IFS in GNIRS for Gemini, Dubbeldam
et al. 2000).

IFSs are becoming increasingly more powerful because (1) more and more
fibers/lenslets can be used, due in turn to technical improvements on
the fibers/lenslets themselves, and to larger CCDs; (2) fibers can be
more closely packed, thus allowing better sampling; (3) IFSs now start
to be implemented on 8-m class telescopes; and (4) IFSs will be coupled
to adaptive optics units, thus allowing for spectral imaging at much
higher spatial resolutions (e.g. Bacon et al. 2001b).

\section{Kinematic information on starbursts and AGN}

Kinematic observations can reveal important clues on dynamical processes
related to the overall behaviour of the host galaxy, to the fueling of
the central regions by gas from the disk, and/or to dynamical feedback,
where material is expulsed by the AGN or by massive newly formed
stars. Some of the most important of these processes are outlined in
this Section for objects with different classes of activity in the
central kpc.

\subsection{Bulges and black holes}

The dynamical study of bulges in general is outside the scope of this
review, but the relationships between the properties of bulges and those
of the massive black holes (MBHs) they may harbour is of interest
here. Full details of the progress in this area can be found in the
comprehensive review by Merritt \& Ferrarese (these proceedings,
p. 000). Kinematic measurements are used to estimate the amount of
enclosed mass, in turn leading to a determination of MBH mass, and to
study the dynamics of bulges. Although other techniques are used (e.g.,
imaging and modeling of central cusps in the light distribution, e.g.,
van der Marel 1999; so-called reverberation mapping of the broad
emission line region of AGN, e.g., Peterson 1993; or direct measurement
of proper motions of individual stars near the nucleus of our Milky Way,
e.g., Genzel et al. 2000), spectroscopy is the most widely used tool to
determine enclosed masses, by measuring the circular rotation velocity
at small radii and deriving the enclosed mass needed to maintain this
rotation (e.g., Miyoshi et al. 1995; Hughes et al., these proceedings,
p. 000). In order to constrain the MBH mass, the rotation measurement
must be made as far inside the bulge of the galaxy as possible, hence
the need for high spatial resolution spectroscopy, e.g., with VLBI or
the {\it HST}.

Kinematic measurements of observable properties of bulges, such as their
velocity dispersion, have recently led to a number of most interesting
correlations between such properties and MBH masses, as described in
detail by Merritt \& Ferrarese (these proceedings).

\subsection{AGN and their hosts}

Kinematic observations of AGN and their host galaxies are used to
investigate important problems, such as how AGN are
fuelled by gaseous material from the disk, or how the energy released by
an AGN influences its surroundings (the first of these is relevant to
all kinds of central activity, not just non-stellar).

Already for two decades, a number of authors have argued on the basis of
theory and modeling that bars, or sets of nested bars, can transport
gaseous material from the disks of galaxies toward their central
regions, basically because part of the angular momentum of the gas is
dissipated by the bars (e.g., Schwarz 1981; Combes \& Gerin 1985;
Shlosman, Frank \& Begelman 1989; Friedli \& Benz 1993). This idea is
hard to confirm kinematically because (1) not all the inflowing gas may
reach the central region of the galaxy (e.g., Knapen et al. 1995a;
Regan, Vogel \& Teuben 1997), (2) observationally one is always
restricted to measuring the line-of-sight velocities, and biased to
observing those around the minor axis, and (3) the gas masses needed to
fuel and sustain even the most powerful AGN or starburst are relatively
small, of the order of a few solar masses per year, and may be brought
about by inflow speeds that are so low that they are hard to measure
(Kenney 1994). As a result, only few authors have published inflow
rates, calculated indirectly by estimating the gravitational torque
exerted on the gas by the stellar bar potential (Quillen et al. 1995),
or by comparison with hydrodynamical models (Regan et al. 1997). Net
inflow velocities of $10-20$~km/s were found, with net gas inflow of
order one to a few solar masses per year. Wong \& Blitz (2000) find some
evidence for inflow from an analysis of \hi\ data of NGC~4736 , but warn
that their inferred inflow velocities are too high and would need
further analysis. Even when considering these few determinations of
inflow velocities, it must be kept in mind that in all cases the
line-of-sight velocity observations are compared to a dynamical
model, which implies a number of assumptions on, e.g., the mass-to-light
ratio or the pattern speed. We conclude that although the overall case
for gas inflow regulated by galactic bars is strong, it remains
difficult to find direct evidence for the inflow.


\subsection{Starburst galaxies}

A causal link has long since been established between the presence of
central starbursts and bars in galaxies (e.g., Heckman 1980; Balzano
1983; Hawarden et al. 1986; Devereux 1987; Puxley, Hawarden \& Mountain
1988; Kennicutt 1994), which, when combined with our knowledge of bars
from theory and modeling, makes for compelling evidence that bars fuel
the starbursts by channeling disk material inwards. This is different
from the situation in AGN hosts, where only recently evidence has been
found for a slight excess of bars in Seyferts when compared to
non-Seyfert galaxies (see Knapen, Shlosman \& Peletier 2000b; Laine et
al. 2001a; also Laine et al., these proceedings, p. 000). As discussed
above, though, the kinematics of this bar-driven inflow of gas remain
difficult to observe directly.

Massive stars in starbursts deposit energy in the surrounding
interstellar medium through supernovae and stellar winds, which give
rise to outflows of gas on galactic scales, so-called superwinds (see
Heckman 2001 for a recent review). Superwinds are ubiquitous in the most
actively star-forming galaxies, both locally (e.g., Lehnert \& Heckman
1996) and at high redshift (Pettini et al. 2001). Depending on the
amount of energy and duration of the wind, and on the properties of the
halo, the superwind may break out of the disk of its host galaxy, and
form a weakly collimated bipolar outflow cone at scales of tens of
kiloparsec, roughly perpendicular to the disk. X-ray emission is
observed from the hot material in the superwind (e.g., Strickland et
al. 2000). Multi-wavelength observations of the cold, warm and hot
phases of the interstellar medium (ISM) associated with the
starburst-driven outflow can yield information on the energetics,
dynamics, and propagation of the starburst, while the dynamical
timescale as determined from the size and kinematics of the outflow can
constrain the starburst age (e.g., Elmegreen 1992; Elmegreen \& Lada
1977; Jogee, Kenney \& Smith 1998; Calzetti et al. 1999; Chapman et
al. 2000a).

\subsection{Circumnuclear star-forming regions}

Nuclear rings are rather common in barred galaxies (although precise
statistics are still missing), and are believed to occur where inflowing
gas slows down near the location of one or more Inner Lindblad
Resonances (ILRs; see review by Shlosman 1999, Sect.~4). Kinematic
observations can be used in this respect to study the gas motions in the
bar, which give information on the bar dynamics and possibly the inflow
velocities and masses (see above). In a first approximation, and in the
case of very weak bars {\it only}, rotation curves can be used to
estimate where the ILRs are expected to lie (basically near the turnover
in the rotation curve).

\begin{figure}
\plotone{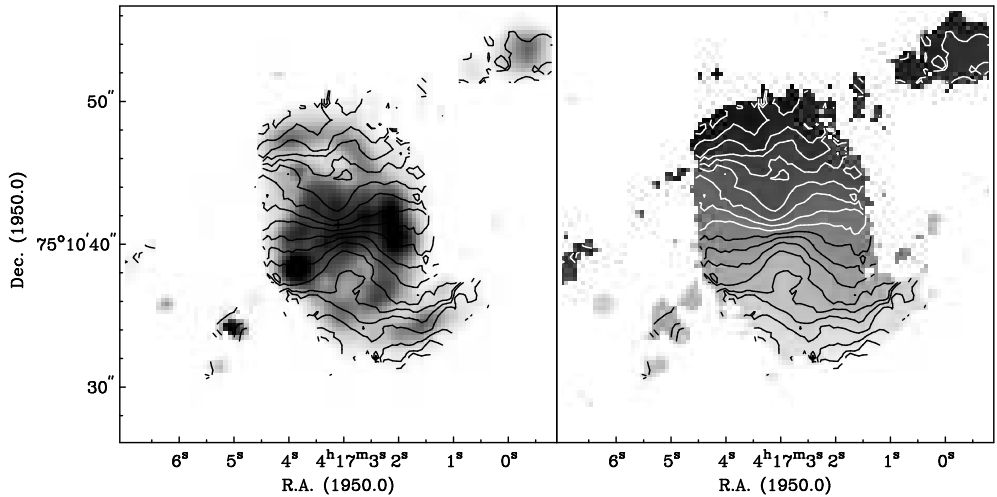}
\caption{H$\alpha$ Fabry-P\'erot total intensity map (left, in greys)
and velocity field (right in greys; in both panels in contours) of the
central region of NGC~1530, as obtained with the 4.2m  
WHT on La Palma. The spatial resolution is $\sim\secd 0.9$.}
\end{figure}

Velocity fields of the circumnuclear regions usually show predominantly
circular rotation, but often also deviations from axisymmetry caused by
non-circular motions. The latter give detailed kinematic information on
spiral arms or inner bars, which can be used to test and constrain
dynamical modelling (e.g. Knapen et al. 2000a; Sect.~4). As an
illustration, we show here TAURUS Fabry-P\'erot results on the core
region of NGC~1530 (Fig.~3), where the rotation is revealed as a spider
diagram outlined by the isovelocity contours in the velocity field. The
wiggles and curves in the contours are due to gas streaming around the
bar, and to spiral density wave streaming in the nuclear star-forming
region (Rela\~no et al., in preparation).

\section{Central kpc kinematics in M100}

M100 (=NGC~4321) has a bar of moderate strength which gives rise to a
particularly clear resonant circumnuclear structure (Knapen et
al. 1995a). It is a prime example of a class of barred galaxies which
host circumnuclear ring-like star-forming or starburst activity (see
review on rings by Buta \& Combes 1996). In the NIR, the morphology of
the core is that of an annular star-forming zone surrounding a small,
nuclear, stellar bar. In the optical, and especially in star formation
tracers such as \ha, individual star-forming sites are clearly
recognised within the ring-like region, and are spread along miniature
spiral armlets. The dust lane morphology shows that these spiral arms
form a two-fold symmetrical, or grand-design, pattern, which connects
through the bar and out into the disk of the galaxy (Knapen et
al. 1995a,b). We modeled these and other observed features in M100
numerically in terms of a resonant structure driven by one bar. The bar
is dissected by the ring-like star-forming region, located between a
pair of ILRs. Although the hydrodynamical modeling and interpretation
agreed in great detail to the morphology as determined from a variety of
images, we decided to test the model results directly with kinematic
measurements.

We used the TAURUS~II instrument in Fabry-P\'erot mode on the WHT on La
Palma to make two-dimensional kinematic observations in the \ha\ line of
the circumnuclear region of M100. The $\sim\secd 0.7$ seeing was well
sampled with $\secd 0.28$ pixels.  After wavelength and phase
calibration we produced a cube containing a series of spatial maps at
increasing wavelength, and thus velocity (Knapen et al. 2000a).

\begin{figure}
\plotone{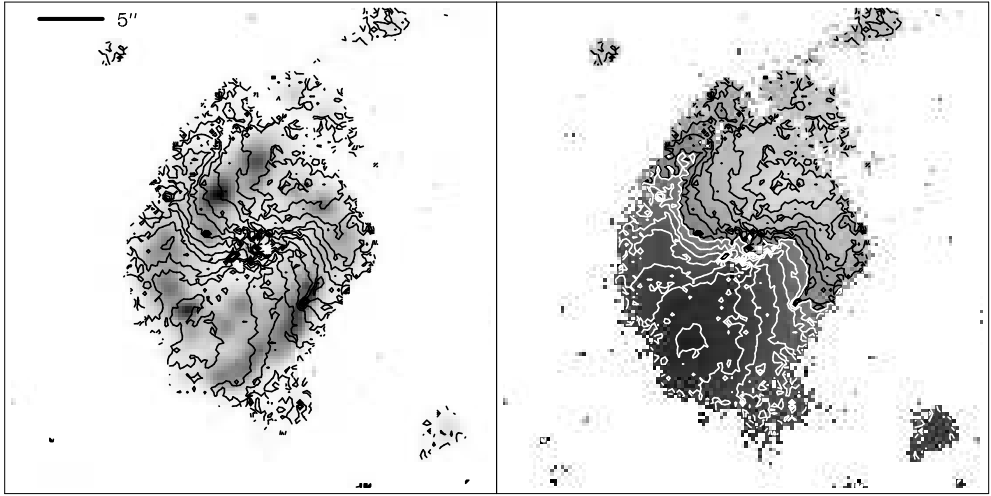}
\caption{H$\alpha$ FP moment maps of the central region of M100: the velocity
field is shown in contours and overlaid on the total intensity H$\alpha$
image (left),
and on the velocity field itself (right). Contours are
separated by 15 \kms, from 1480\kms\ to 1675\kms. Black contours in
the right panel are low velocities; the first white contour is at 1585\kms.
N is up, E to the left.}
\end{figure}

After determining which channels of the data set were free of \ha\ line
emission, we used those channels to determine and subtract the continuum
emission. A moment analysis was used to produce a total intensity map
(moment zero) and a velocity field (moment one) in \ha; these are shown
in Fig.~4. Proof of the reliability of the FP technique is the fact that
the total intensity \ha\ map, thus reconstructed, is comparable in
quality to the narrow-band \ha\ image published by Knapen et
al. (1995b), with an estimated spatial resolution of \secd 0.6 -- \secd
0.7$\!$. Although the velocity field is dominated by circular motions,
and a rotation curve can be derived, it does show important deviations
from circular motion.  The positions where these deviations take place
can be related to the total intensity map. A more detailed analysis of
the data cube confirms that the deviations are due to streaming motions
in the spiral arms, and to a lesser extent, to gas streaming along the
inner part of the bar (see Knapen et al. 2000a for more details). We
thus showed kinematically that the spiral armlets are density wave
spiral arms, and that the elongation in the NIR isophotes is in fact due
to a bar, as reported before (Knapen et al. 1995a,b).

\begin{figure}
\plotone{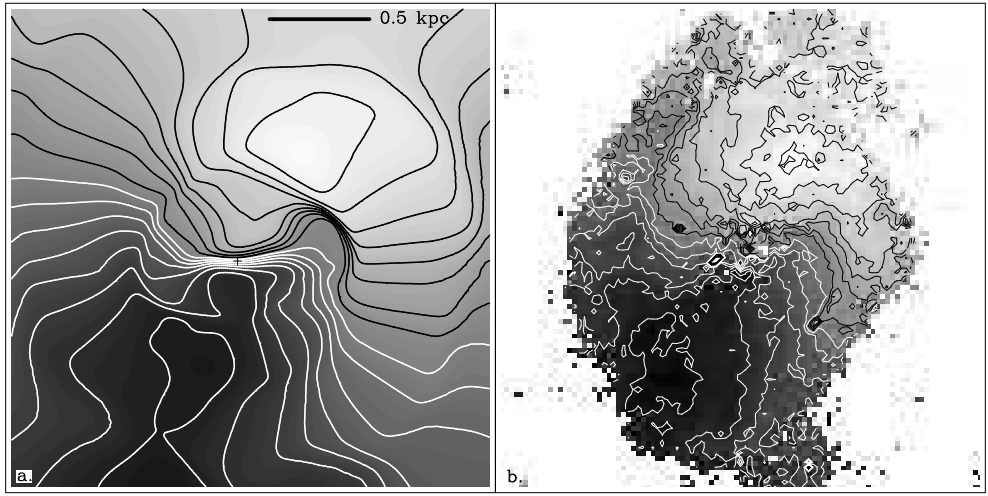}
\caption{Left panel ({\it a.}): Gas velocity field as derived from the
numerical model of Knapen et al. (1995a). Contour separation is 15\kms, and
the scale is indicated in the top right hand corner. 
N is up, E to the right. The position angle of the major axis is
as in M100. Resolution is comparable to our \ha\ data. Right panel ({\it
b.}): For comparison,  \ha\ velocity field of Fig.~3 at
the same scale and orientation, and with the same contour
separation. From Knapen et al. (2000a).}
\end{figure}

To make a detailed kinematic comparison with the new \ha\ FP data, we
produced a velocity field from our SPH dynamical model of the CNR of M100
(Knapen et al. 1995a), which was made before the kinematic data became
available. The result of that comparison is shown in Fig.~5. The
qualitative and quantitative agreement between the model and \ha\
velocity fields, as outlined in this Figure, gives further support to
our interpretation of the CNR in terms of a resonance region driven by
one bar, dissected by the ring-like region which houses the star-forming
spiral armlets.

\section{The core of NGC 5248}

Nuclear spirals have recently been discovered in a considerable number
of spiral galaxies of varying types, thanks exclusively to the
availability of high-resolution imaging, especially in the NIR, both
from space using the {\it HST} and from the ground using adaptive optics
(e.g., Ford et al. 1994; Phillips et al. 1996; Grillmair et al. 1997;
Devereux, Ford \& Jacoby 1997; Dopita et al. 1997; Carollo, Stiavelli \&
Mack 1998; Elmegreen et al. 1998; Malkan, Gorjian \& Tam 1998; Rouan et
al. 1998; Laine et al. 1999; Martini \& Pogge 1999; Regan \& Mulchaey
1999; Chapman, Morris \& Walker 2000b; Tran et al. 2001). This nuclear
spiral structure, most often seen in dust, occurs at scales of a few
tens to hundreds of parsec, and is in the great majority of cases
flocculent in appearance. In a handful of galaxies, however, nuclear
grand-design spirals are found (NGC~5248 in Laine et al. 1999; UGC 12138
and NGC 7682 in Martini \& Pogge 1999). The importance of nuclear
spirals lies in a number of key areas like the nature of spiral
structure, its behavior in the presence of dynamical resonances, or the
possible role of spirals in fuelling nuclear activity.

There are three main theoretical schemes to explain the existence of
nuclear spiral arms within the location of the ILR(s). It is relatively
easy to distinguish between these schemes with suitable observations,
{\it if} the right observations at sufficiently high spatial resolution
are available. First, a nuclear spiral pattern can exist within the
inner ILR of the outer spiral pattern. The nuclear spiral must in this
case have a much higher pattern speed than the dynamically independent
large spiral pattern in the disk (Heller \& Shlosman 1994; Englmaier \&
Shlosman 2000). Second, as described and modelled by Englmaier \&
Shlosman (2000), the large-scale bar or outer spiral pattern can drive
the nuclear spiral, and the pattern speeds of both sets of spirals must
be equal. The potential well must in this case be shallow enough to
prevent the spiral from winding up and destroying itself by
dissipation. A third possibility is that nuclear spiral arms, but
exclusively flocculent ones, are formed by acoustic instabilities, as
described by Elmegreen et al. (1998, see also Montenegro, Yuan \&
Elmegreen 1999).

A good illustration of the power of a complete kinematic analysis of the
central regions of galaxies is given by the case study of the nuclear
spiral arms in the core of NGC~5248, which is in fact the first
kinematic study of nuclear spirals. The spiral structure in NGC~5248 can
be identified as very faint outer stellar arms, at scales of tens of
kpc, classical optically bright star formation arms at scales of a few
kpc, gas/dust arms at the inner ends of the star formation spirals, and
the grand-design nuclear spiral at scales of tens to a few hundreds of
pc, reported by Laine et al. (1999) on the basis of adaptive optics (AO)
NIR imaging. In addition, NGC~5248 hosts a circumnuclear star-forming
ring, containing a number of UV-bright knots of star formation (see Maoz
et al. 2001). In addition to this ring with a diameter of some 14 arcsec
(1.1~kpc), there is another pseudo-ring, visible in \ha, which encircles
the nucleus at a radius of 1 arcsec. This innermost ring thus lies in
the same region as the grand-design nuclear spirals, but its relation to
those spirals, or in fact its origin, is not known (Laine et al. 2001;
Maoz et al. 2001).

\begin{figure}
\plotfiddle{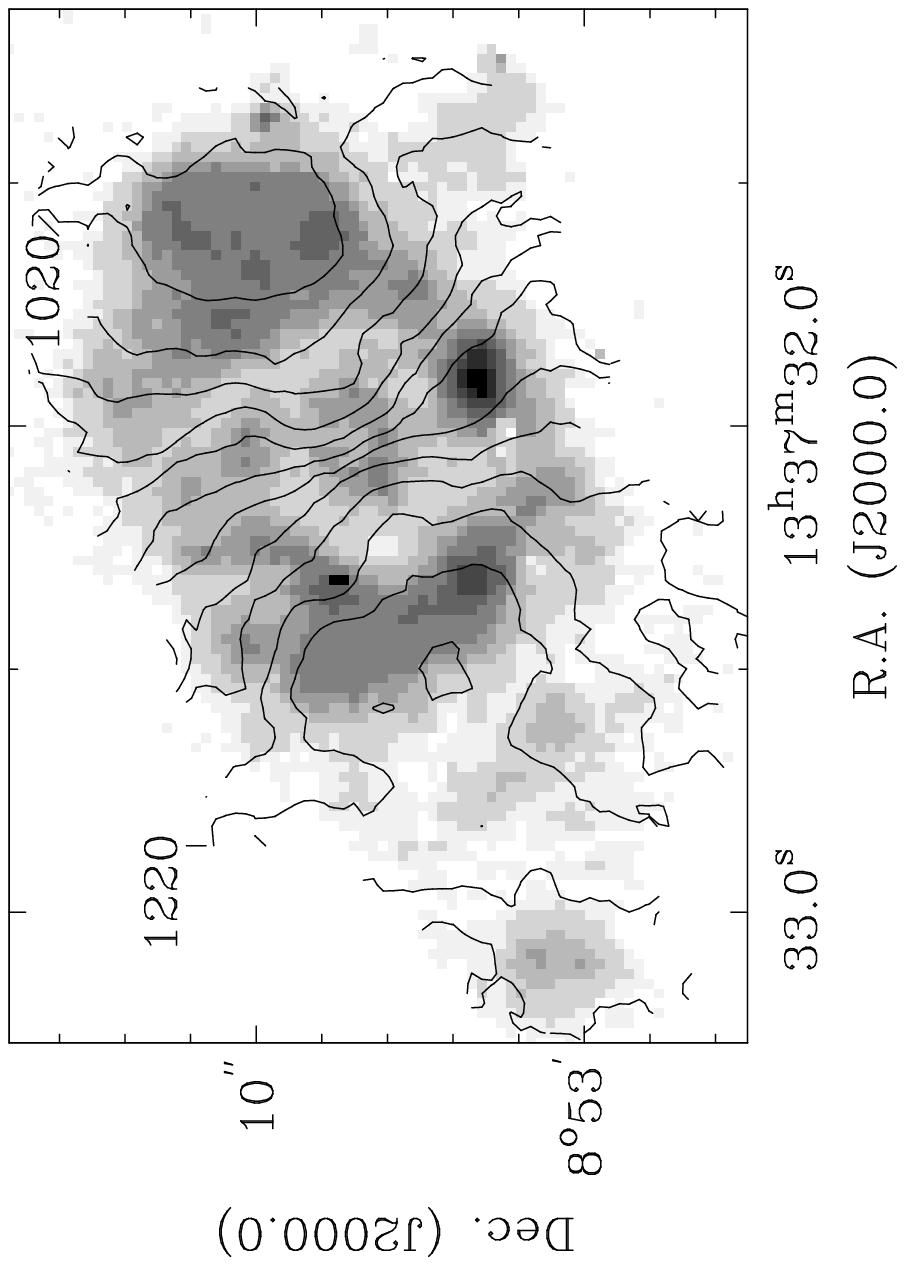}{6cm}{-90}{90}{90}{-220}{500}
\caption{H$\alpha$ Fabry-P\'erot total intensity map and velocity field
of the central region of NGC~5248, as obtained with the WHT. Spatial 
resolution $\sim\secd 0.8$, area shown $\sim36\min$
across, and selected contours are labeled with velocity in
km/s. Reproduced with permission from Laine et al. (2001).}


\vspace{2.5cm}

\plotfiddle{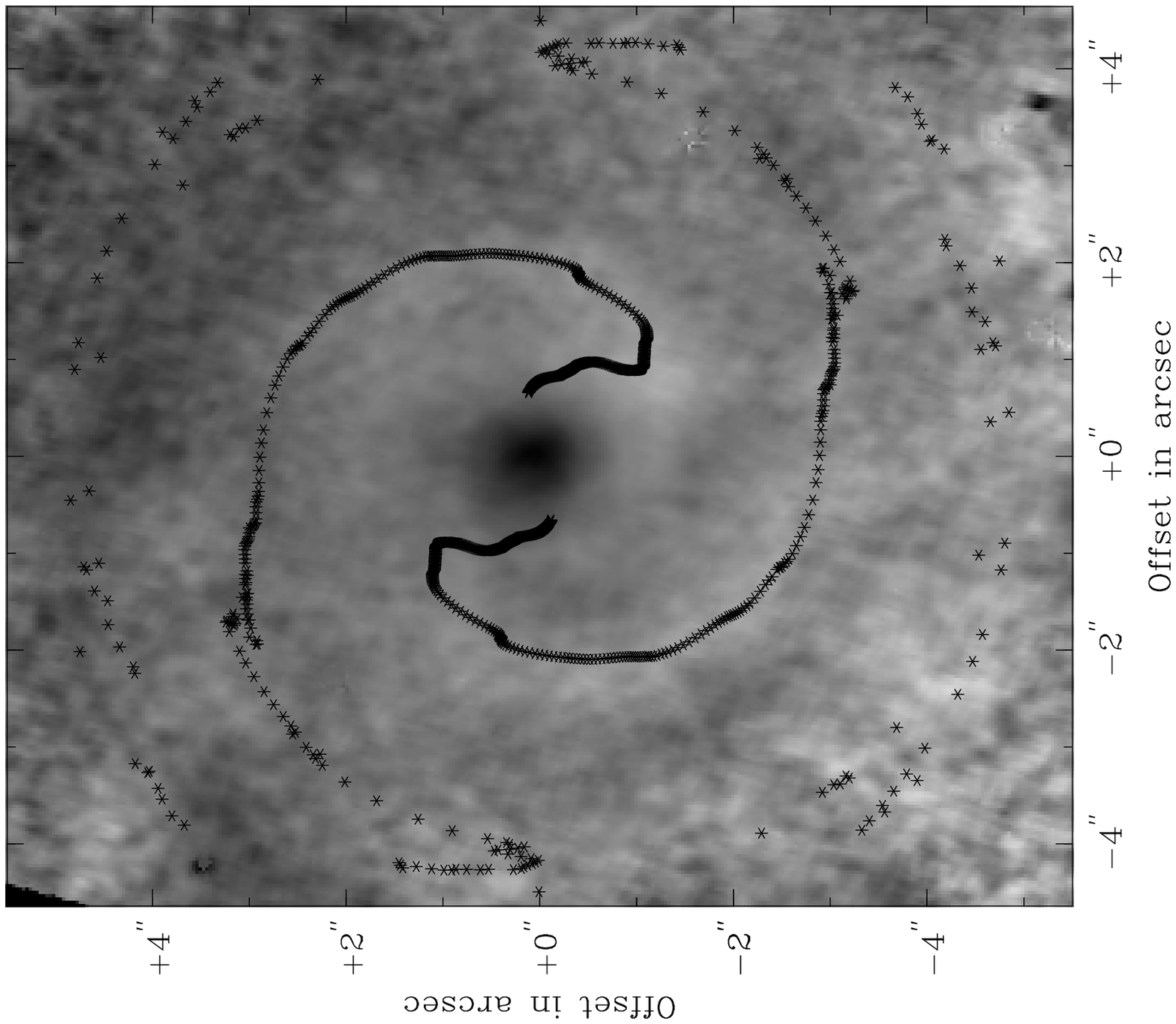}{4.7cm}{-90}{45}{45}{-250}{225}
\plotfiddle{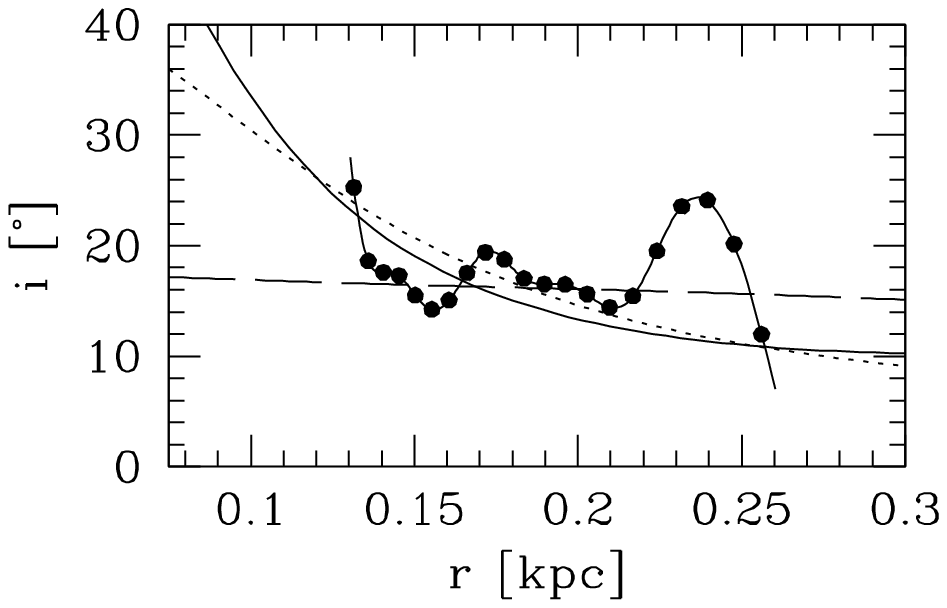}{0cm}{0}{60}{60}{-80}{-320}
\caption{Fitted nuclear spirals from AO $J-K$ image (left), and the
observed pitch angle compared to models (right). Reproduced with
permission from Laine et al. (2001).}
\end{figure}

Laine et al. (1999) suggested that the nuclear spiral may have formed
through the second of the three mechanisms described above, i.e., it is
dynamically coupled to the outer spiral. In order to test this
hypothesis, we obtained a fully sampled two-dimensional set of kinematic
data using TAURUS on the WHT. The resulting data set has a spatial
resolution of around 0.9 arcsec (68~pc) and a spectral resolution
corresponding to 18.9~km/s (see Laine et al. 2001 for details). The
morphology is dominated by the nuclear ring with its strong star
formation knots, but also the innermost pseudo-ring is visible. The
kinematic structure is dominated by circular rotation, where the closed
contours on both sides of the nucleus imply that the rotation is falling
after reaching a local maximum (Fig.~6). In marked contrast to the
circumnuclear region in M100 (see above), which is morphologically
similar, non-circular motions are practically absent from the velocity
field of the core of NGC~5248 -- from the circumnuclear as well as from
the nuclear spiral region. The well-resolved rotation curve as derived
from the velocity field (Fig.~6) shows a relatively shallow rise in the
curve (in contrast to, e.g., M100 [Knapen et al. 2000a]; or many other
spiral galaxies [Sofue \& Rubin 2001]).

In order to consider the main parameter determining the nature of the
nuclear spiral pattern, namely its pattern speed in comparison with that
of the main spiral pattern, a combination of morphology, kinematics and
modelling is needed. Laine et al. (2001) measured the pitch angle of the
nuclear spiral arm in the $J-K$ AO image from Laine et al. (1999).  They
also derived the rotation curve from the kinematic velocity field in the
same radial range (\secd 1.8 -- \secd 3.4). Using analytical modelling
based upon Englmaier \& Shlosman (2000) and the rotation curve, the
pitch angle of the nuclear spiral was subsequently modeled
(Fig.~7). From this comparison Laine et al. concluded that the observed
(from the $J-K$ AO image) and modeled (from the rotation curve) pitch
angles agree well for a low pattern speed ($\Omega_{\rm
P}=13~$km\,s$^{-1}$\,kpc$^{-1}$) and ``normal'' ISM sound speed ($c_{\rm
s}=9-16$ km\,s$^{-1}$). This is the pattern speed derived by Patsis,
Grosb{\o}l \& Hiotelis (1997) for the outer spiral, and the best
agreement in the modelling is thus found by using equal, low, pattern
speeds for the outer and nuclear spirals. High pattern speed values for
the nuclear spiral are specifically excluded because they are
incompatible with the observed arm pitch angle. The acoustic spiral
theory of Elmegreen et al. (1998) is not applicable in this case of a
grand-design two-armed spiral, because that mechanism will produce
chaotic and multi-armed spirals.

The conclusion from the kinematic observations of Laine et al. (2001),
after combining them with high-resolution imaging and modelling, is thus
that the nuclear spiral rotates at the same rate as the outer spiral
(the second of the three theoretical scenarios outlined above). This
implies that the spiral structure is dynamically coupled from scales of
a few tens of pc to several kpcs. Transport and distribution of gas from
the disk to the core region, over orders of magniitude in scale, is
controlled by coupled mechanisms. 

\section{Final remarks and future prospects}

A considerable variety of techniques is used to study the
kinematics of gas and stars at different wavelengths, and at different
spatial scales, in the central regions of galaxies. We have reviewed
some of these, and indicated how they can increase our understanding
of various aspects of galaxies and especially those hosting
(circum)nuclear starbursts and AGN. Although all of the techniques
mentioned will remain in use in the future, some have particular
promise. First, integral field spectrographs will become more powerful
(a) when more and more image elements can be fed to spectrographs and onto
ever larger detectors, thus increasing the spatial resolution, sampling, and/or
the field of view, (b) when used on 8-m class telescopes, and (c) when used in
conjunction with AO units. Secondly, upgrades to existing
millimeter interferometers will continue to increase the spatial
resolution and sensitivity of kinematic data obtained, mostly in the
CO lines. ALMA is expected to revolutionise also this field when it
becomes operational, sometime later this decade, allowing observations at
spectacularly improved sensitivities and spatial resolutions of the
central kpc regions of galaxies with a wide variety of central
activity. Observations with such improved or new facilities, along
with continued progress in modeling and theory, will no doubt lead to
fascinating new insights into the how, why and when of central
activity in galaxies.

In a number of examples we have shown that there is firm evidence that
the central kpc regions of galaxies are firmly embedded in, and
dynamically directly coupled to the spiral arms in the main disk of the
galaxy. In M100, a combination of extensive imaging and kinematic
observations at various wavelengths with detailed numerical modelling
has shown that the circumnuclear structure in this galaxy is resonant in
origin, and is being maintained and fed by a bar structure. Within the
ring-like region of enhanced star formation and gas density, which is
related to the position of a pair of ILRs, density wave spiral arms
occur, directly coupled dynamically to the spiral arms in the main disk
of the galaxy. In NGC~5248, the dynamical coupling of spiral arm systems
is occurring at an even larger range of length scales. Kinematic
observations in conjunction with analytic and numerical modeling and
adaptive optics imaging shows that the nuclear grand-design spiral at
scales of tens of pc rotates with equal pattern speed to the outer
spiral, at scales of a few kpc.

{\it Acknowledgements} I thank my collaborators on the various projects
mentioned in this paper, especially Seppo Laine, Isaac Shlosman, and
Reynier Peletier and Shardha Jogee. I thank Harald Kuntschner and the
SAURON team for their help in preparing Figure~2. The William Herschel
Telescope is operated on the island of La Palma by the Isaac Newton
Group in the Spanish Observatorio del Roque de los Muchachos of the
Instituto de Astrof\'\i sica de Canarias.

\end{document}